\documentstyle[twocolumn,epsfig,prl,aps]{revtex}
\begin{document}
\newcommand{\lbl}[1]{\label{eq:#1}}
\newcommand{ \rf}[1]{(\ref{eq:#1})}
\newcommand{\be}{\begin{equation}}
\newcommand{\en}{\end{equation}}
\newcommand{\bea}{\begin{eqnarray}}
\newcommand{\ena}{\end{eqnarray}}
\newcommand{\lapprox}{%
\mathrel{%
\setbox0=\hbox{$<$}
\raise0.6ex\copy0\kern-\wd0
\lower0.65ex\hbox{$\sim$}
}}
\newcommand{\gapprox}{%
\mathrel{%
\setbox0=\hbox{$>$}
\raise0.6ex\copy0\kern-\wd0
\lower0.65ex\hbox{$\sim$}
}}

\def\mpi{M_\pi}
\def\fpi{F_\pi}
\def\mhat{{\widehat m}}

\newcommand{\tr}{\mbox{\rm tr}}

\begin{titlepage}
\begin{flushright}
CPT-99/P.3870\\
UAB-FT-469\\
\end{flushright}

\begin{center}
\begin{bf}

{\Large \bf Decay of Pseudoscalars  into Lepton Pairs\\[0.3cm]
 and Large--$N_C$ QCD} \\[2cm]
\end{bf}
{\large M. Knecht$^a$, S. Peris$^b$, M. Perrottet$^a$ and E. de Rafael$^a$}
\\[1cm]
$^a$ Centre de Physique Th{\'e}orique, CNRS--Luminy, Case 907\\ 
F-13288 Marseille Cedex 9, France.\\[.1cm]
$^b$ Grup de F{\'\i}sica T{\`e}orica and IFAE, Universitat Aut{\`o}noma de
Barcelona\\
 E-08193 Barcelona, Spain. \\[3cm]
{\bf Keywords:} \begin{minipage}[t]{9.5cm} Chiral Symmetry,
Chiral Perturbation Theory, Electroweak Theory, Large--$N_C$ QCD.
\end{minipage}
\end{center}

\vfill
\begin{abstract}
The counterterm combination that describes the decay of pseudoscalar 
mesons into charged lepton pairs at lowest order in chiral 
perturbation theory is considered within the framework of QCD in the limit of 
a large number of colours $N_C$. When further restricted to the lowest meson 
dominance approximation to large-$N_C$ QCD, our results agree 
well with the available experimental data.
\end{abstract}
\vfill

\end{titlepage}

\begin{titlepage}

{\mbox{  }}

\end{titlepage}


\wideabs{
\title{Decay of Pseudoscalars  into Lepton Pairs
and Large--$N_C$ QCD }
\author{
M. Knecht$^a$, S. Peris$^b$, M. Perrottet$^a$ and E. de Rafael$^a$
}

\address{
$^a$ Centre de Physique Th{\'e}orique, CNRS--Luminy, Case 907,
F-13288 Marseille Cedex 9, France\\
$^b$ Grup de F{\'\i}sica T{\`e}orica and IFAE, Universitat Aut{\`o}noma de
Barcelona, E-08193 Barcelona, Spain}
\maketitle

\begin{abstract}
The counterterm combination that describes the decay of pseudoscalar 
mesons into charged lepton pairs at lowest order in chiral 
perturbation theory is considered within the framework of QCD in the limit of 
a large number of colours $N_C$. When further restricted to the lowest meson 
dominance approximation to large-$N_C$ QCD , our results agree 
well with the available experimental data.
\end{abstract}

}

{\bf 1.}~The theoretical study of the $\pi^0$ and $\eta$ decaying into lepton 
pairs and
the comparison with the experimental rates~\cite{AH99,PDG98}
offers an interesting possibility to test our understanding of the 
long--distance 
dynamics of the Standard Model. These processes are dominated by the exchange 
of two virtual photons and it is therefore phenomenologically useful to 
consider the branching ratios normalized to the two-photon branching ratio 
($P=\pi^0,\eta$)
\bea
\!\!\!\!\!\!\!\!
R(P\to\ell^+\ell^-) &=& \frac{Br (P\to\ell^+\ell^-)}{Br (P\to\gamma\gamma)}
\nonumber\\ 
&=&
2\bigg(\frac{\alpha m_{\ell}}{\pi M_P}\bigg)^2\,
\beta_{\ell}(M_P^2)\,\vert{\cal A}(M_P^2)\vert^2,\lbl{width}
\ena
with $\beta_{\ell}(s) = \sqrt{1-4m_{\ell}^2/s}$. The unknown dynamics
is then contained in the amplitude ${\cal A}(M_P^2)$.
To lowest order in the chiral expansion the contribution to this
amplitude arises from the two graphs of Fig.~1 with the result      
\be
{\cal A}(s) = \chi_P(\mu)
+\frac{N_C}{3}\bigg[\,-\frac{5}{2}+ 
\frac{3}{2}\ln\bigg(\frac{m_{\ell}^2}{\mu^2}\bigg)\,+\,C(s)\,\bigg],
\lbl{amp}
\en
where $\chi_{\pi^0}=\chi_{\eta}=-(\chi_1+\chi_2)/4
\equiv\chi$, with $\chi_1$ and $\chi_2$ the couplings of the two counterterms
which describe the direct  interactions of pseudoscalar mesons with lepton
pairs to lowest order in the chiral expansion~\cite{SLW92}
\bea
&&
{\cal L}_{P\ell^+\ell^-}= \frac{3i}{32}\left(\frac{\alpha}{\pi}\right)^2
\,{\bar \ell}\gamma^\mu\gamma_5 \ell
\nonumber\\
&&\times
\,\big[\chi_1\tr(Q_RQ_RD_\mu UU^{\dagger}
-Q_LQ_LD_\mu U^{\dagger}U)
\nonumber\\
&&
+\,
\chi_2\tr(U^{\dagger}Q_RD_{\mu}UQ_L-UQ_LD_{\mu}U^{\dagger}Q_R)
\big].\lbl{CTlag}
\ena
Here the unitary matrix $U$ describes the meson fields and 
$Q_L =Q_R =\mbox{\rm diagonal}(2/3,-1/3,-1/3)$.
The function $C(s)$ in Eq.~\rf{amp} corresponds to a finite 
three--point loop integral which
can be expressed in terms of the dilogarithm function 
${\mbox{Li}}_2(x)\,=\,-\int_0^x(dt/t)\ln(1-t)$. 
For $s<0$, its expression reads

\bea
C(s) &=& \frac{1}{\beta_{\ell}(s)}
\bigg[\,{\mbox{Li}}_2\bigg( 
\frac{\beta_{\ell}(s) -1}{\beta_{\ell}(s) +1}
\bigg)\,+
\,\frac{\pi^2}{3}\lbl{Csub}\nonumber\\
&&
+\,\frac{1}{4}\ln^2\bigg(
\frac{\beta_{\ell}(s)-1}{\beta_{\ell}(s)+1}
\bigg)\,\bigg].
\ena

\begin{figure}[t]
\centerline{\psfig{figure=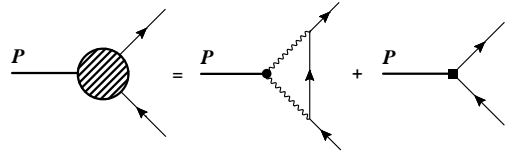,height=3.0cm}}
\caption[The lowest order contributions]
{The lowest order contributions to the 
$P\to\ell^+\ell^-$ decay amplitude. The second graph denotes the 
contribution from the counterterm lagrangian of Eq.~\rf{CTlag}.}
\end{figure}

The corresponding expression for $s>4m_{\ell}^2$ is obtained by analytic 
continuation, using the usual $i\epsilon$ prescription.
The loop diagram of Fig.~1 
originates 
from the usual coupling of the light pseudoscalar mesons to a photon pair 
given by the well--known Wess--Zumino anomaly~\cite{WZ}. The divergence 
associated with this diagram has been 
renormalized within the ${\overline{\mbox{MS}}}$ minimal subtraction scheme 
of dimensional regularization. The logarithmic dependence on the 
renormalization scale $\mu$ 
displayed in the above expression is compensated by the scale dependence of 
the combination $\chi(\mu)$ of renormalized low--energy constants defined 
above.
Let us stress here that, as shown explicitly in Eq.~\rf{amp} and in contrast 
with the usual situation in the purely mesonic sector, this scale dependence 
is not suppressed in the large--$N_C$ limit, since it does not arise from 
meson loops. The evaluation of $\chi(\mu)$ will be 
the main subject of this paper.

It has recently been shown~\cite{GDP98} that, when evaluated within the 
chiral $U(3)\otimes U(3)$ framework and in the $1/N_C$ expansion, the 
$\vert\Delta S\vert = 1$ $K_L^0\to\ell^+\ell^-$ 
transitions can also be described by the expressions \rf{width} and \rf{amp}, 
with an effective constant 
$\chi_{K^0_L}$ containing an additional piece
from the short--distance contributions~\cite{BF97}. Of course, a cast-iron
understanding of these transitions is very important\cite{K0L} as  
the evaluation of
$\chi(\mu)$ could then have a potential impact on possible constraints
on physics beyond the Standard Model. We comment on this issue at the end 
of the paper.




{\bf 2.}~As a first step towards its subsequent evaluation
we shall identify the coupling constant $\chi$ in 
terms of a QCD correlation function. For that purpose, consider the matrix 
element 
 of the light quark isovector 
pseudoscalar density 
$P^3(x)\,=\,{1\over 2}({\bar u}i\gamma_5 u - {\bar d}i\gamma_5 d)(x)$ 
between leptonic states in the chiral limit.
In the absence of weak interactions, and to lowest non--trivial 
order in the fine structure constant, this matrix element is given by
the integral
\bea
&&\!\!\!\!
<\ell^-(p')\,\vert\,P^3(0)\,\vert\,\ell^-(p)>\qquad\nonumber\\
&&\quad
=e^4\int \frac{d^4q}{(2\pi)^4}\,
\frac{{\bar u}(p')\gamma^\mu[{\not\! p}'- {\not\! q} + m_{\ell}]
\gamma^\nu u(p)}{[(p'-q)^2-m_{\ell}^2]q^2(p'-p-q)^2}\nonumber
\\
&&\quad\times
\,i\int d^4x\int d^4y\,e^{iq\cdot x}e^{i(p'-p-q)\cdot y}\nonumber\\
&&\quad\times
<0\,\vert\,T\{j_\mu^{\mbox{\scriptsize{em}}}(x)
              j_\nu^{\mbox{\scriptsize{em}}}(y)P^3(0)\}\,\vert\,0>,
\lbl{matel}
\ena
with 
$j_\mu^{\mbox{\scriptsize{em}}}={1\over 3}
(2{\bar u}\gamma_\mu u - {\bar d}\gamma_\mu d - {\bar s}\gamma_\mu s)$.
In the chiral limit, the QCD three--point correlator appearing in this 
expression is an order parameter of spontaneous chiral symmetry breaking. 
This ensures that it has a smooth behaviour at short distances. 
In particular, 
Bose symmetry and parity conservation of the strong interactions yield
\bea
&&\!\!\!\!\!\!\!\!\!\!
\int d^4x\int d^4y\,e^{iq_1\cdot x}e^{iq_2\cdot y}
\nonumber\\
&&\times
<0\,\vert\,T\{j_\mu^{\mbox{\scriptsize{em}}}(x)
              j_\nu^{\mbox{\scriptsize{em}}}(y)P^3(0)\}\,\vert\,0>\nonumber\\
&&
=\ \frac{2}{3}\,\epsilon_{\mu\nu\alpha\beta}q_1^\alpha q_2^\beta
\,{\cal H}(q_1^2,q_2^2,(q_1+q_2)^2),\lbl{VVP}
\ena
with ${\cal H}(q_1^2,q_2^2,(q_1+q_2)^2) = {\cal H}(q_2^2,q_1^2,(q_1+q_2)^2)$. 
For very large (euclidian) 
momenta, the leading short--distance behaviour of this correlation function 
is given by
\bea
&&\!\!\!\!
\lim_{\lambda\to\infty}\,{\cal H}
\big((\lambda q_1)^2,(\lambda q_2)^2,(\lambda q_1 + \lambda q_2)^2\big)
\nonumber\\
&&
=
\ -\frac{1}{2\lambda^4}<{\bar \psi}\psi>\,
\frac{q_1^2+q_2^2+(q_1+q_2)^2}{q_1^2q_2^2(q_1+q_2)^2}\nonumber\\
&&
\ +
\,{\cal O}\bigg(\frac{\alpha_s}{\lambda^4},\frac{1}{\lambda^6}\bigg).
\lbl{high}
\ena
Actually, what matters for the convergence of the integral in Eq.~\rf{matel} 
is the leading short--distance singularity of the $T-$product of the two 
electromagnetic currents, which corresponds to
\bea
&&\!\!\!\!
\lim_{\lambda\to\infty}\,{\cal H}
\big((\lambda q)^2,(p'-p-\lambda q)^2,(p'-p)^2\big)
\nonumber\\
&&
=
\ -\frac{1}{\lambda^2}<{\bar \psi}\psi>\,
\frac{1}{q^2(p'-p)^2}
\, +\,
{\cal O}\bigg(\frac{\alpha_s}{\lambda^2},\frac{1}{\lambda^3}\bigg),
\lbl{high2}
\ena
and which implies that the loop integral in Eq.~\rf{matel} is indeed 
convergent. The QCD corrections
of order ${\cal O}(\alpha_s)$ in Eqs.~\rf{high} and \rf{high2} will not be 
considered here. 
Let us 
however notice that since the pseudoscalar density $P^3(x)$ and the 
single--flavour $<{\bar \psi}\psi>$ condensate 
share the same anomalous dimension, the power--like 
fall--off displayed by Eqs.~\rf{high} and \rf{high2} is canonical, i.e. it is 
not modified by powers of logarithms of the momenta.

On the other hand, at very low momentum transfers, the same correlator can be 
computed within 
Chiral Perturbation Theory (ChPT). At lowest order, it is saturated by 
the pion--pole contribution, given by the anomalous coupling of a neutral pion,
emitted by the pseudoscalar source $P^3(0)$, 
to the two electromagnetic currents, i.e.
\be
{\cal H}(0,0,(q_1+q_2)^2)\ =\ \,\frac{N_C}{8\pi^2}\,\frac{<{\bar \psi}\psi>}
{F_0^2}\,\frac{1}{(q_1+q_2)^2}\ +\cdots\ ,\lbl{chiral}
\en
where the ellipsis stands for higher orders in the low--momentum expansion 
and where $F_0$ denotes the pion decay constant in the chiral limit.
The matrix element $<\ell^-(p')\,\vert\,P^3(0)\,\vert\,\ell^-(p)>$ itself 
may also be evaluated in ChPT. At lowest order, it is given by the diagrams 
of Fig.~1, where the (off--shell) pion is now emitted by the pseudoscalar 
source $P^3(0)$. The result reads, with $t=(p'-p)^2$,
\bea
&&\!\!\!\!\!\!\!\!
<\ell^-(p')\,\vert\,P^3(0)\,\vert\,\ell^-(p)>
\big\vert_{\mbox{~\scriptsize{ChPT}}}\\
&&
=\  - \frac{ie^4}{32\pi^4t}\,
\frac{<{\bar \psi}\psi>}{F_0^2}\,m_{\ell}{\bar u}(p')\gamma_5u(p)
\,{\cal A}(t),\nonumber
\ena
with the function ${\cal A}(t)$ defined in Eqs.~\rf{amp} and \rf{Csub}.
The contribution of the loop diagram of Fig.~1 is obtained upon replacing,
in Eq.~\rf{matel}, the three--point QCD correlator by its 
lowest order chiral expression given in Eq.~\rf{chiral}.
The coupling constant $\chi(\mu)$ is thus 
given by the residue of the pole at $t=0$ of the matrix element 
$<\ell^-(p')\,\vert\,P^3(0)\,\vert\,\ell^-(p)>$, after subtraction of the 
contribution of the two--photon loop, i.e.
\bea
&&
\frac{\chi(\mu)}{32\pi^4}
\,\frac{<{\bar \psi}\psi>}{F_0^2}\,m_{\ell}{\bar u}(p')\gamma_5 u(p)
=\ -\frac{2i}{3}\,{\bar u}(p')\gamma_{\lambda}\gamma_5 u(p)
\nonumber\\
&&\times
\lim_{(p'-p)^2\to 0}\,\int \frac{d^dq}{(2\pi)^d}
\frac{(p'-p)^2}{[(p'-q)^2-m_{\ell}^2]q^2(p'-p-q)^2}\nonumber\\
&&\times\,
(p'-p-2q)_{\alpha}\bigg[\,q^{\alpha}(p'-p)^{\lambda}
-\,(p'-p)^{\alpha}q^{\lambda}
\,\bigg]\nonumber\\
&&\times\,
\bigg[\,{\cal H}(q^2,(p'-p-q)^2,(p'-p)^2)\,
-\,{\cal H}(0,0,(p'-p)^2)\,\bigg]\lbl{exact}.\nonumber\\
\ena
Since the integral occurring in the above expression diverges, we have 
regularized it by analytical continuation in the space--time dimension $d$. 
The coupling $\chi(\mu)$ on the left--hand side is then 
defined by the ${\overline{\mbox{MS}}}$ minimal subtraction prescription, as 
in Eq.~\rf{amp}. 
Keeping only the contributions that do not vanish as  
$(p'- p)^2$ goes to zero, and neglecting terms of order 
${\cal O}(m_{\ell}^2/\Lambda_H^2)$, where $\Lambda_H\sim M_{\rho}$ 
is a typical hadronic scale for non--Goldstone mesonic states, 
we obtain a somewhat simpler expression,
\bea
&&\!\!\!\!
\frac{\chi(\mu)}{32\pi^4}
\,\frac{<{\bar \psi}\psi>}{F_0^2}
=\ -\bigg(1\,-\,\frac{1}{d}\bigg)\,\int \frac{d^dq}{(2\pi)^d}\,
\bigg(\frac{1}{q^2}\bigg)^2\nonumber\\
&&\!\!\!
\times\lim_{(p'- p)^2\to 0}\,
(p'-p)^2\bigg[\,{\cal H}(q^2,q^2,(p'-p)^2)\nonumber\\
&&\qquad\qquad\qquad\qquad\quad
-\,
{\cal H}(0,0,(p'-p)^2)\,\bigg].\lbl{chisimp}
\ena
In order to perform this integral, one needs to extend the knowledge of the 
three--point correlation 
function ${\cal H}(q_1^2,q_2^2,(q_1+q_2)^2)$ in the chiral limit beyond its 
behaviour at energies very high, 
Eq.~\rf{high}, or at energies very low, Eq.~\rf{chiral}. 
Stated like that, in full generality,
this represents a rather formidable task. As we shall next
show, it is possible, however, following the examples 
discussed recently in Refs.~\cite{KPdeR98,KPdeR99}, to proceed further 
within the
framework of the $1/N_{C}$--expansion in QCD~\cite{tH}.

{\bf 3.}~In the limit where
the number of colours
$N_C$ becomes infinite, with $\alpha_{s}\times N_{C}$ staying finite,
%
%
%
%
the QCD spectrum reduces to an infinite tower of zero--width 
mesonic resonances~\cite{W79}, and the leading large--$N_C$ contributions to 
the three--point correlator \rf{VVP} are given by the tree--level 
exchanges of these resonances in the various channels, as shown in Fig.~2.
This involves couplings of the resonances 
among themselves and to the external sources which, just like the masses 
of the resonances themselves, cannot be fixed in the absence of an
explicit solution of QCD in the large--$N_C$ limit. 
In this limit, however, the
analytical structure of the three--point function in Eq.~\rf{VVP} is very
simple:~the  singularities in each channel consist of a succession of
{\it simple poles}.
Furthermore, the quantity appearing in Eq.~\rf{chisimp} has the general 
structure
\bea
&&\!\!\!\!
\lim_{(p'- p)^2\to 0}\,
(p'-p)^2\,{\cal H}(q^2,q^2,(p'-p)^2)\ =
\ -\frac{1}{2}\frac{<{\bar \psi}\psi>}{F_0^2}
\nonumber\\
&&\qquad\times
\sum_V\,M_V^2\,\bigg[\,\frac{a_V}{(q^2-M_V^2)}\,-
\,\frac{b_V q^2}{(q^2-M_V^2)^2}\,\bigg]\lbl{gen}\ ,
\lbl{largeN}
\ena
where {\it a priori} the sum extends over the {\it infinite} 
spectrum of vector 
resonances of QCD in the large--$N_C$ limit. Equation \rf{largeN} follows 
from 
the fact that its left--hand side enjoys some additional properties:~i)~In 
the pseudoscalar channel, only the pion pole 
survives, while massive pseudoscalar resonances cannot contribute. ii)~The 
momentum transfer in the two vector channels is the same. iii)~Its 
high--energy behaviour is fixed by Eq.~\rf{high2}.

%

\begin{figure}[t]
\centerline{\psfig{figure=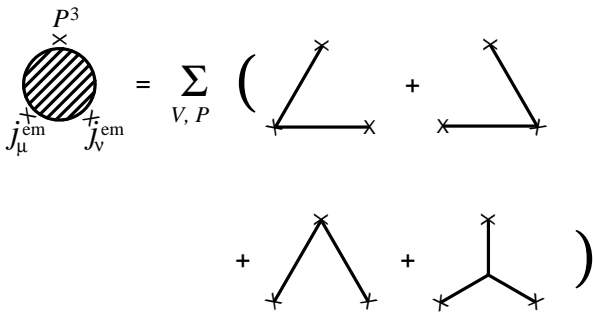,height=5.0cm}}
\caption[Large-Nc]
{The con\-tri\-bu\-tions to the ve\-ctor--\-vec\-tor--\-pseu\-do\-sca\-lar 
three--\-point func\-tion in the large--\-$N_C$ limit of QCD. The sum extends 
over the infinite number of zero-width vector ($V$) and pseudoscalar ($P$) 
states.
}
\end{figure}

\noindent
Even though the constants $a_V$ and $b_V$ 
depend on the masses and couplings of the vector resonances in an unknown 
manner, they are however constrained by the two conditions 
\be
\sum_V\,a_V = \frac{N_C}{4\pi^2}
, \quad 
\sum_V\,(a_V\,-\,b_V)\,M_V^2 = 2F_0^2,\lbl{cond}
\en
which follow from Eqs.~\rf{chiral} and \rf{high2}, respectively. Notice that 
there are no contributions from the perturbative QCD continuum to these sums.
Taking the first of these conditions 
(which, coming from the anomaly, has no ${\cal O}(\alpha_s)$ corrections) 
into account, we obtain
\be
\chi(\mu) = \frac{5N_C}{12}
-2\pi^2\,\sum_V\,
\bigg[\,a_V\ln\left(\frac{M_V^2}{\mu^2}\right)\,-\,b_V\,\bigg].
\lbl{mainAppr}
\en
This equation, together with the two conditions \rf{cond}, constitutes 
the central result of our paper. This is as far as the large--$N_C$ limit 
allows us to go. Let us point out 
that the scale dependence of $\chi(\mu)$ is correctly 
reproduced by the expression \rf{mainAppr}, again as a consequence of the 
first condition in Eq.~\rf{cond}.
However, in order to obtain a numerical estimate of $\chi(\mu)$ additional 
assumptions are needed.




{\bf 4.}~In order to proceed further, we shall consider the 
Lowest Meson Dominance (LMD) approximation to the large--$N_C$ 
spectrum of vector meson resonances discussed in~\cite{PPdeR98}. This 
approximation has been shown to reproduce very well the relevant 
low--energy
constants of the ${\cal O}(p^4)$ chiral Lagrangian~\cite{GL84} and the
electromagnetic $\pi^+ -\pi^0$ mass difference~\cite{KPdeR98}. 
In our case, it corresponds
to the assumption that the sums occurring  in Eqs.~\rf{gen} and
\rf{cond} are saturated by the lowest lying vector meson octet. 
In the LMD approximation to large--$N_C$ QCD, the two conditions \rf{cond} 
completely pin down the 
two quantities $a_V$ and $b_V$ in terms of $F_0$ and of the mass $M_V$ of 
this lowest lying vector meson octet,
\be
a_V^{\mbox{\scriptsize LMD}}=\frac{N_C}{4\pi^2}\quad\mbox{and}
\quad b_V^{\mbox{\scriptsize LMD}}=\frac{N_C}{4\pi^2}\,-
\,\frac{2F_0^2}{M_V^2}.\lbl{abLMD}
\en
In fact, within the LMD approximation of large--$N_C$ QCD, it is easy to 
write down the expression for the correlation function 
${\cal H}(q_1^2,q_2^2,(q_1+q_2)^2)$ which 
correctly interpolates between the high energy behaviour in Eq.~\rf{high} 
and the 
ChPT result in Eq.~\rf{chiral}~\cite{fnte2}

\bea
&&
{\cal H}^{\mbox{\scriptsize LMD}}(q_1^2,q_2^2,(q_1+q_2)^2)\ =\ 
-\frac{1}{2}<{\bar \psi}\psi>\nonumber\\
&&\times
\frac{q_1^2+q_2^2+(q_1+q_2)^2-M_V^4a_V^{\mbox{\scriptsize LMD}}/F_0^2}
{(q_1^2-M_V^2)(q_2^2-M_V^2)(q_1+q_2)^2}.
\lbl{HvLMD}
\ena
Notice that this expression also correctly reproduces the behaviour in 
Eq.~\rf{high2}.
With the results of Eq.~\rf{abLMD}, and for $N_C$=3, it follows from 
Eq.~\rf{mainAppr} that
\be
\chi^{\mbox{\scriptsize LMD}}(\mu) =
\frac{11}{4}\,-\,\frac{3}{2}
\ln\left(\frac{M_V^2}{\mu^2}\right)
-4\pi^2\frac{F_0^2}{M_V^2}.
\lbl{chiLMD}
\en
Numerically, using the physical values $F_0=92.4$~MeV and 
$M_V=M_{\rho}=770$~MeV, we obtain
\be
\chi^{\mbox{\scriptsize LMD}}(\mu = M_V)
\ =\ 2.2\pm 0.9,\lbl{numLMD}
\en
where we have allowed for a systematic theoretical error of 40\%, as a rule 
of thumb 
estimate of the uncertainties attached to the large--$N_C$ and LMD 
approximations.
The predicted ratios of branching ratios in
Eq.~\rf{width} which follow from this result \cite{fnte1} 
are displayed in Table \ref{table1}.
We conclude that, within errors, the LMD--approxi\-ma\-tion to 
large--$N_C$ QCD
re\-pro\-duces well the observed rates of pseudoscalar mesons decaying into 
lepton pairs.

{\bf 5.} At present, the most accurate experimental determination of
the $K_L^0\to\mu^+\mu^-$ branching ratio~[20] gives the
result: 
$Br(K_L^0\to\mu^+\mu^-)=(7.18\pm 0.17)\times 10^{-9}$. In the framework of the
$1/N_C$ expansion and using the experimental branching ratio~[2] 
$Br(K_L^0\to\gamma\gamma)=(5.92\pm 0.15)\times 10^{-4}$, this 
leads to a unique solution for an
{\it effective} $\chi_{K_L^0}=5.17\pm 1.13$. Furthermore, following Ref.~[5],  
$\chi_{K_L^0}=\chi - {\cal N}\ \delta \chi_{SD}$ where 
${\cal N}=(3.6/g_8 c_{\rm red})$  normalizes the 
$K_L^0\to\gamma\gamma$ amplitude. The coupling $g_8$ governs the $\Delta I=1/2$
rule, the constant $c_{\rm red}$ is defined in Ref.~[5] and  
$\delta \chi_{SD}^{\mbox{\tiny\rm Standard}}=(+1.8\pm 0.6)$ is the short
distance contribution in the Standard Model~[6]. 

Therefore our understanding of the {\it short distance}  
contribution to this process completely hinges on our understanding of the
{\it long distance} constant ${\cal N}$ and therefore of the $\Delta I=1/2$
rule in the $1/N_C$ expansion. Moreover, $c_{\rm red}$ is regretfully 
very unstable in the chiral and large-$N_C$
limits, a behaviour that surely points 
towards the need to have higher
order corrections under control. For instance, for $M_{\pi}=0, M_{K}\not=0,
N_C\rightarrow \infty$ one obtains $c_{\rm red}=0$, while for 
$M_{\pi}= M_{K}=0, N_C\rightarrow \infty$ (and the external $K_L^0$ off shell) 
one obtains $c_{\rm red}=-4/3$ instead. The analysis of Ref.~[5] 
uses $c_{\rm red}\simeq +1$ and $g_8\simeq 3.6$,
where these numbers are obtained phenomenologically by requiring agreement 
with  the two-photon decay of $K_L^0, \pi^0, \eta$ and $\eta'$ as well 
as $K\to 2\pi, 3\pi$. Should we use these values of $c_{\rm red}$ and $g_8$ 
and Eq. (19) we would obtain $\chi_{K_L^0}=0.4\pm 1.1$, 
corresponding to a ratio 
$R(K_L^0\to\mu^+ \mu^-)=(2.24\pm 0.41)\times 10^{-5}$ which is $2.5\sigma$ 
above the experimental value 
$R(K_L^0\to\mu^+ \mu^-)=(1.21\pm 0.04)\times 10^{-5}$. 

\begin{table}[b]
\caption[Results]{The values for the ratios $R(P\to\ell^+\ell^-)$ obtained 
within the LMD approximation to large--$N_C$ QCD and the comparison with 
available experimental results.}
\begin{tabular}{ccc}
$R$ & LMD & Experiment\\ \hline
$R(\pi^0\to e^+e^-)\times 10^{8}$ & $6.2\pm 0.3$ & $7.13\pm 0.55$~\cite{AH99}
  \\ \hline
$R(\eta\to \mu^+\mu^-)\times 10^{5}$ & $1.4\pm 0.2$ & $1.48\pm 0.22
$~\cite{PDG98} \\ \hline
$R(\eta\to e^+e^-)\times 10^{8}$ & $1.15\pm 0.05$ & ---\\
\end{tabular}
\label{table1}
\end{table}

In view of these uncertainties we conclude
that it does not seem to be possible, within our understanding 
of long--distance
effects in the electroweak interactions, to argue that 
$K_L^0\to\mu^+ \mu^-$ is,
at present, a useful decay to constrain physics beyond the Standard
Model.

We thank Ll. Ametller and A. Pich for discussions. 
Work supported in part by TMR, EC--Contract No. 
ERBFMRX--CT980169. The work of S.P. is also supported by AEN98-1093.


\begin{references}
\bibitem{K0L}
         We refer the reader to Ref.~\cite{GDP98} for further details. See 
         also Refs. \cite{dAmb98} and \cite{Val98} for an alternative approach.
\bibitem{fnte1}
        In the case of the 
        decay into a muon pair, $m_{\mu}/\Lambda_H$ 
        corrections lower the result of 
        Eq.~\rf{numLMD} by about 7\%, and are taken into account in the 
        numerical values of Table I.
\bibitem{fnte2}
         A similar analysis for the 
         vec\-tor--\-vec\-tor--\-sca\-lar and 
         vec\-tor--\-axial--\-pseudo\-sca\-lar three--\-point 
         functions can be found in Refs.~\cite{MS94} and \cite{M97}, 
         respectively.
\end{references}
\end{document}